\documentclass[12pt]{article}
\usepackage{amsmath,amsfonts,amssymb,amsthm,bm,cite}

\DeclareMathOperator{\csch}{csch}     

\renewcommand{\b}{\beta}            
\newcommand{\braket}[2]{\langle#1\mathbin|#2\rangle} 
\newcommand{\dl}{\delta}            
\newcommand{\dn}{{\mathord{\downarrow}}} 
\newcommand{\ga}{\gamma}            
\newcommand{\half}{\tfrac{1}{2}}    
\newcommand{\ket}[1]{|#1\rangle}    
\newcommand{\ketbra}[2]{|#1\rangle\langle#2|} 
\newcommand{\la}{\lambda}           
\newcommand{\om}{\omega}            
\newcommand{\set}[1]{\{\,#1\,\}}    
\newcommand{\sg}{\sigma}            
\newcommand{\shalf}{{\scriptstyle\frac{1}{2}}} 
\renewcommand{\th}{\theta}          
\newcommand{\twobytwo}[4]{\begin{pmatrix}#1& #2\\ #3& #4\end{pmatrix}}
\newcommand{\up}{{\mathord{\uparrow}}} 
\newcommand{\vf}{\varphi}           
\newcommand{\vth}{\vartheta}        
\newcommand{\word}[1]{\quad\mbox{#1}\quad} 
\newcommand{\x}{\times}             
\newcommand{\7}{\dagger}            
\renewcommand{\.}{\cdot}            

\newcommand{\gs}{\mathrm{gs}}       
\newcommand{\HF}{\mathrm{HF}}       

\newcommand{\vecform}{\bm}              
\newcommand{\PP}{\vecform{P}}           
\newcommand{\pp}{\vecform{p}}           
\newcommand{\ppp}{\vecform{\pi}}        
\newcommand{\qq}{\vecform{q}}           
\newcommand{\rr}{\vecform{r}}           
\newcommand{\RR}{\vecform{R}}           
\newcommand{\uu}{\vecform{u}}           
\newcommand{\xx}{\vecform{x}}           
\newcommand{\zz}{\vecform{z}}           

\newtheorem{thm}{Theorem}               

\makeatletter
\def\section{\@startsection{section}{1}{\z@}{-3.5ex plus -1ex minus
 -.2ex}{2.3ex plus .2ex}{\large\bfseries}}
\def\subsection{\@startsection{subsection}{2}{\z@}{-3.25ex plus -1ex
 minus -.2ex}{1.5ex plus .2ex}{\normalsize\bfseries}}
\makeatother

\begin{document}

\title{Harmonium as a laboratory
\\
for mathematical chemistry}

\author{Kurusch Ebrahimi-Fard and Jos\'e M. Gracia-Bond\'ia
\\ \\
Departamento de F\'isica Te\'orica, Universidad de Zaragoza,
\\
50009 Zaragoza, Spain}

\date{11 April 2011}

\maketitle

\begin{abstract}
Thanks to an algebraic duality property of reduced states, the Schmidt
best approximation theorems have important corollaries in the rigorous
theory of two-electron moleculae. In turn, the ``harmonium model'' or
``Moshinsky atom'' constitutes a non-trivial laboratory bench for
energy functionals proposed over the years (1964--today), purporting
to recover the full ground state of the system from knowledge of the
reduced 1-body matrix. That model is usually regarded as solvable;
however, some important aspects of it, in particular the exact energy
and full state functionals ---unraveling the ``phase dilemma'' for the
system--- had not been calculated heretofore. The solution is given
here, made plain by working with Wigner quasiprobabilities on phase
space. It allows in principle for thorough discussions of the merits
of several approximate functionals popular in the theoretical chemical
physics literature; in this respect, at the end we focus on Gill's
``Wigner intracule'' method for the correlation energy.
\end{abstract}




\section{Introduction}
\label{sec:introibo}

At the heart of theoretical chemistry sits the Coulson program of
replacing the wave function by the (reduced) 12-variable 2-body
matrix~$\rho_2$~\cite{Coulsondixit}. For this function of twelve
variables the basic energy functional is trivially known.
Unfortunately, the $N$-representability problem for~$\rho_2$ has never
been solved in an usable way. Density functional theory (DFT) was
unleashed by the Hohenberg--Kohn discovery that somehow the same
information as in the wave function or in the 2-matrix is stored in
the humble 3-variable electronic density. However, the energy
functional encoding such information is unknown, and must have
strangely non-local properties ---in relation with the existence of
negative ions, for instance~\cite{libronaranja}. A natural alternative
is the Gilbert program~\cite{nonlocalmonger} based on the 6-variable
1-body matrix~$\rho_1$, for which the $N$-representability question is
solved.%
\footnote{After a long period of stasis, recently there has been
progress on the representability problem for 2-matrices, and on
pending questions of representability for 1-matrices
\cite{NeilBirgitta,ErdahlBane,IdemMan,Klyachko}.}

Now, possession of the 1-body matrix for an $N$-electronic system does
not allow exact inference of the corresponding 2-body matrix~$\rho_2$,
or of the correlation energy. It is natural to diagonalize $\rho_1$
and seek to expand $\rho_2$ in terms of eigenfunctions of~$\rho_1$
(``natural orbitals'') and its eigenvalues $0\le\nu_j\le 1$
(``occupation numbers''), with $\sum_j\nu_j=N$. Over the years,
starting with the work by M\"uller~\cite{ForgottenCity}, approximate
relations between~$\rho_2$ and $\rho_1$ and energy functionals based
on this spectral analysis of~$\rho_1$ have been suggested and tried,
with various results.

Two-electron systems constitute an exception to our ignorance. For
them it is feasible to a large extent to reconstruct $\rho_2$ from
$\rho_1$, and so fulfil the Coulson and Gilbert programs. As we shall
recall, that it may be so stems simply from Schmidt's approximation
theorems~\cite{Erhard}. For a reduced 1-density of the kind
$$
\rho_1(\xx,\xx') = \bigl( \up_1\up_{1'} + \dn_1\dn_{1'} \bigr)
\rho_1(\rr,\rr') = \bigl( \up_1\up_{1'} + \dn_1\dn_{1'} \bigr)
\sum_j n_j\phi_j(\rr)\phi_j^*(\rr'),
$$
with $\sum_jn_j=1$, the corresponding 2-density matrix is given by
\begin{align}
&\rho_2(\xx_1,\xx_2,\xx'_1,\xx'_2) 
= \frac{1}{2}\bigl( \up_1\dn_2 - \dn_1\up_2 \bigr)
\bigl( \up_{1'}\dn_{2'} - \dn_{1'}\up_{2'})
\notag \\
&\quad \x \begin{pmatrix}
\phi_1(\rr_1) & \phi_2(\rr_1) & \phi_3(\rr_1) & \cdots
\end{pmatrix} \begin{pmatrix}
c_1 & & & \\ & c_2 & & \\ & & c_3 & \\ & & & \ddots 
\end{pmatrix} \begin{pmatrix}
\phi_1(\rr_2) \\ \phi_2(\rr_2) \\ \phi_3(\rr_2) \\ \vdots \end{pmatrix}
\label{eq:rho-two} 
\\
&\quad \x \begin{pmatrix}
\phi^*_1(\rr'_1) & \phi^*_2(\rr'_1) & \phi^*_3(\rr'_1) & \cdots 
\end{pmatrix} \begin{pmatrix}
c_1^* & & & \\ & c_2^* & & \\ & & c_3^* & \\ & & & \ddots
\end{pmatrix} \begin{pmatrix}
\phi_1^*(\rr'_2) \\ \phi_2^*(\rr'_1) \\ \phi_3^*(\rr'_2) \\ \vdots
\end{pmatrix}.
\notag
\end{align}
Alas, the recipe, although exact, is underdetermined: of the $c_j$ we
only know that $|c_j|^2=n_j$. This is a ``phase dilemma'' of 1-body
matrix functional theory. We work here only with states described by
real wave functions: this still leaves us with an infinite number of
signs to account~for.

\smallskip

The harmonium model originally proposed by Moshinsky~\cite{Moshinsky}
consists of two spin-$\half$ fermions trapped in a harmonic potential
and repelling each other with a Hooke's law force, as well. It
constitutes an analogue to a two-electron atom, helpful to illustrate
the main ideas of reduced density matrices and correlation energy in
an exactly solvable context. Its ground state, as well as the reduced
density matrix and the pair distribution, exhibiting correlation, are
well known in the standard wave function formalism~\cite{Davidson}.
Recently, this model has been recruited for suggesting approximate
Ans\"atze for correlation energy density~\cite{MarchCCA}, and other 
purposes~\cite{GiveUp}.

Nevertheless, within the standard formalism it is not at all obvious
how to go about the sign conundrum. We manage here to sidestep all the
difficulties, and exemplify the two-electron state theorem for
harmonium, by working with Wigner functions on phase space instead.
Mathematically, Wigner phase space states are equivalent to density
matrices, sporting of course the same number of variables. However,
our approach calls for a different type of physical intuition,
encapsulating some of the non-locality which is the bane of~DFT.

Like one-body density matrix functional theory, Wigner functional
theory sits midway between ordinary~DFT and the Coulson proposal.
Currently Gill and coworkers~\cite{Gillito2} pursue a vigorous program
to understand correlation energies via Wigner intracules: our method
also sits transversally midway between the 6-variable Gilbert and Gill
programs. We mention as well a different fusion of the Schmidt and
Wigner ways, related to information theory~\cite{LuzanoNoAnadaluz},
and a recent quantum phase space view of harmonium by Dahl
\cite{DahlMosh}.

\smallskip

The plan of the article: we start by recalling the classical
contribution by Schmidt, with a pedagogical bent. We show how to
derive from it the exact energy functional for two-electron systems,
determined up to an unknown configuration of signs. After that, we
describe harmonium by phase space methods; these provide the fast lane
towards an elegant and complete description of this system. Finally,
we examine Gill's Wigner intracule method for correlation energy from
our results on harmonium.

\section{Schmidt's theorem and its meaning for heliumalikes}
\label{sec:vea}

Little more than singular value decomposition of matrices is involved
here. Let $\Psi(x_1,x_2)$ be the wave function describing a pure state
of any system with two electrons ---including the helioid series,
starting with $\mathrm{H}^-,\mathrm{He}$, and molecules like
$\mathrm{H}_2$. For simplicity, and because in absence of magnetic
fields this is actually the case for such systems, we assume that
$\Psi$ is real. Let us expand~$\Psi$ in suitable orthonormal basis
(ONB), also taken real,
\begin{align*}
\Psi(x_1, x_2) &= \sum_{ij}\braket{f_ih_j}\Psi f_i(x_1)h_j(x_2) =:
\sum_{ij}f_i(x_1)\,c_{ij}\,h_j(x_2)
\\
&:= \begin{pmatrix} f_1(x_1) & f_2(x_1) & f_3(x_1) & \cdots
\end{pmatrix} \, C \, \begin{pmatrix} h_1(x_2) \\ h_2(x_2) \\ h_3(x_2)
\\ \vdots \end{pmatrix}.
\end{align*}
Suppose the eigenvalue problem for the (real positive symmetric)
matrix $CC^t$ is solved by
$$
CC^tu_j = \sg^2_ju_j.
$$
We order the singular values: $\sg_1\ge\sg_2\ge\sg_3\ge\cdots\ge0$.
Then the $u_j$ are the columns of an orthogonal matrix $U$. Define~now
$v_i=C^tu_i/\sg_i$. These are at least partially the columns of
another orthogonal matrix, say~$V$, since
$$
v_i^tv_j = \frac{\braket{u_i}{CC^tu_j}}{\sg_i\sg_j} =
\frac{\sg^2_j\braket{u_i}{u_j}}{\sg_i\sg_j} = \dl_{ij}.
$$
Add to them any orthonormal basis for $\ker C$. Notice moreover that
$$
u_i^tCv_j = \frac1{\sg_i}u_i^tCC^tu_j = \dl_{ij}\sg_i, \word{or} U^tCV
= \begin{pmatrix} \sg_1 &&& \\ & \sg_2 && \\ && \sg_3 & \\
&&& \ddots \end{pmatrix} =: \Sigma.
$$
We just have reached the singular value decomposition of (real)
matrices:
\begin{equation}
C = U\Sigma V^t,
\label{eq:ultima-ratio}
\end{equation}
with $\Sigma$ diagonal nonnegative and $U,V$ orthogonal, whose columns
are ``singular vectors'', respectively left-hand and right-hand ---see
for instance~\cite[Chap.~5]{Meyer}. Notice the non-unicity of~$V$.%
\footnote{This decomposition amounts to the canonical ``Schmidt
series'' for an operator and the ``Schmidt decomposition" of tensor
products, currently taught to students interested in entanglement,
decoherence and quantum information theories~\cite{caballerosos}.}

\smallskip

Before proceeding, we remark that Schmidt in~\cite{Erhard} showed that
the \textit{best} finite-rank approximation (in the sense of any
orthogonally invariant norm) to~$C$ is given simply by
$$
C^{(k)} = U^{(k)}\Sigma^{(k)}{V^{(k)}}^t,
$$
where $\Sigma^{(k)}$ is obtained just by cutting out $\sg_{k+1}=
\sg_{k+2}=\cdots=0$, and $U^{(k)},V^{(k)}$ are the matrices obtained
by conformally keeping in $U,V$ the first $k$~singular vectors. Most
important for us, we note as well that $CC^t$ (save the conventional
factor~2) precisely represents the 1-body density associated
to~$\Psi$:
\begin{align*}
\int\Psi(x, x_2)\Psi(x', x_2)\,dx_2 &= \sum_{ijk}f_i(x)\,c_{ij}c_{kj}\,
f_k(x') = \sum_l\sg_l^2\phi_l(x)\phi_l(x')
\\
&=: \sum_l(\nu_l/2)\phi_l(x)\phi_l(x') =: \half\rho_1(x, x');
\end{align*}
where we have made the spectral expansion of that density: the $\nu_l$
are the chemists' \textit{occupation numbers} and the $\phi_l$ the
\textit{natural orbitals}. We witness here a wonderful convergence of
mathematical relevance and physical import. Schmidt approximation then
implies that a state made out of $K(K-1)/2$ configurations can be
rewritten with $K$~configurations.

\smallskip

Going back to~\eqref{eq:ultima-ratio} and denoting:
$$
V^t\begin{pmatrix} h_1 \\ h_2\\ \vdots \end{pmatrix} = \begin{pmatrix}
\vf_1 \\ \vf_2 \\ \vdots \end{pmatrix},
$$
it is clear that $C^tC$ is the matrix representation of $\rho_1$ in the
ONB $\set{\ket{h_j}}$ and $CC^t$ the one in the ONB $\set{\ket{f_i}}$,
and that
\begin{align*}
\Psi(x_1, x_2) &= \sum_i\pm\sqrt{\nu_i/2}\,\phi_i(x_1)\vf_i(x_2)
\\
&=: \sum_i\pm\sqrt{\nu_i/2}\,\phi_i(x_1)\int dx\,\phi_i(x)\Psi(x, x_2).
\end{align*}
The sign indetermination at this stage cannot be removed. The
challenge is to identify the right configuration of signs for each
system from energy considerations.

Since moreover $\Psi$ is skewsymmetric, $\vf_i$ must be a natural
orbital itself. For instance, for the singlet ground state of the
helioids,
$$
\Psi_\mathrm{gs}(x_1, x_2) = \frac1{\sqrt2}\bigl(\up_1\dn_2 -
\dn_1\up_2\bigr)\Psi(\rr_1, \rr_2),
$$
if we choose
\begin{align*}
\phi_{2i-1}(x) &= \up\phi_i(\rr), \quad \phi_{2i}(x) = \dn\phi_i(\rr),
\\
\word{then} \vf_{2i-1}(x) &= \dn\phi_i(\rr), \quad
\vf_{2i}(x) = -\up\phi_i(\rr).
\end{align*}
For the triplet states we have:
\begin{align*}
\Psi_\mathrm{t1}(x_1, x_2) &= \up_1\up_2\sum_id_i[\ga_i(\rr_1)\om_i(\rr_2)
- \om_i(\rr_1)\ga_i(\rr_2)];
\\
\Psi_\mathrm{t0}(x_1, x_2) &= \frac1{\sqrt2}\bigl(\up_1\dn_2 +
\dn_1\up_2\bigr)\sum_id_i[\ga_i(\rr_1)\om_i(\rr_2) -
\om_i(\rr_1)\ga_i(\rr_2)];
\\
\Psi_\mathrm{t-1}(x_1, x_2) &= \dn_1\dn_2\sum_id_i[\ga_i(\rr_1)\om_i(\rr_2)
- \om_i(\rr_1)\ga_i(\rr_2)];
\end{align*}
so, if we choose
\begin{align*}
\phi^1_{2i-1}(x) &= \up\ga_i(\rr), \quad
\phi^1_{2i}(x) = \up\om_i(\rr),
\\
\word{then} \vf^1_{2i-1}(x) &= \up\om_i(\rr), \quad
\vf^1_{2i}(x) = -\up\ga_i(\rr);
\end{align*}
and similarly in the other cases.

\smallskip

The Schmidt theorem has a far-reaching generalization for best
approximations of $N$-body wave functions in terms of orbitals of
complementary reduced density matrices. A good place to see that in
action is~\cite{Davidson}.

\section{Harmonium on phase space}
\label{sec:y-oiga}

The Hamiltonian for harmonium in Hartree units is
\begin{equation}
H = \frac{p_1^2}{2} + \frac{p_2^2}{2} + \frac{k}{2}(r_1^2 + r_2^2)
- \frac{\dl}{4} r^2_{12}.
\label{eq:Mosh-atom}
\end{equation}
Introduce extracule and intracule coordinates, respectively given by
$$
\RR = (\rr_1 + \rr_2)/\sqrt{2}, \qquad \rr = (\rr_1 - \rr_2)/\sqrt{2},
$$
with conjugate momenta
$$
\PP = (\pp_1 + \pp_2)/\sqrt{2}, \qquad \pp = (\pp_1 - \pp_2)/\sqrt{2}.
$$
Therefore
$$
H = H_R + H_r = \frac{P^2}{2} + \frac{\om^2 R^2}{2} + \frac{p^2}{2}
+ \frac{\mu^2r^2}{2}.
$$
We have introduced the frequencies $\om = \sqrt k$ and $\mu =
\sqrt{k-\dl}$; for both particles to remain in the potential well,
assume $\dl< k$.

\smallskip

The translation to phase space of the analysis in the previous section
was accomplished in~\cite{Pluto}. One could depart somewhat from the
treatment in this reference, though. There, following in particular
Springborg and Dahl~\cite{SD87}, by outright transfer from the density
matrix formalism, we regard the 1-body and 2-body Wigner functions
defined below as respectively $2\x2,4\x4$ matrices. It is however more
natural, as advised by Wigner himself~\cite{ColericoEnCPark} and
effected in~\cite{Miranda}, to describe spin by combining row and
column components to have simpler transformation properties; these are
none other than the ones corresponding to the tensor representations
of the rotation group. In particular for the reduced 1-density for a
two-electron system in a singlet spin state, the spin factor:
$$
\frac{1}{2}\bigl( \up_1\up_{1'} + \dn_1\dn_{1'} \bigr)
$$
is just the identity scalar, and can be omitted. In general, for the
1-body Wigner function one organizes the spin factors in terms of the
Cartesian components of the spin operator vector.

We may concentrate on the spinless part of the ground state
henceforth. In general, the spinless Wigner quasiprobability
(normalized to one) corresponding to a (real) 2-particle wave function
$\Psi$ is given by
\begin{align}
&P_\Psi(\rr_1,\rr_2;\pp_1,\pp_2)
\label{eq:phase-two}
\\
&\; = \frac{1}{\pi^6} \int \rho_2(\rr_1 - \zz_1, \rr_2 - \zz_2; \rr_1
+ \zz_1, \rr_2 + \zz_2)\, e^{2i(\pp_1\.\zz_1 + \pp_2\.\zz_2)} \,d^3z_1
\,d^3z_2,
\notag
\end{align}
with $\rho_2(\rr_1,\rr_2;\rr'_1,\rr'_2)=\Psi(\rr_1, \rr_2)\Psi(\rr'_1,
\rr'_2)$. The definition extends to transition matrices $\ketbra{\Phi}
{\Phi'}$ as well:
\begin{align*}
&P_{\Phi\Phi'}(\rr_1,\rr_2;\pp_1,\pp_2)
\\
&\; = \frac{1}{\pi^6} \int \Phi(\rr_1 - \zz_1, \rr_2 - \zz_2)
\Phi'(\rr_1 + \zz_1, \rr_2 + \zz_2) \,e^{2i(\pp_1\.\zz_1 + \pp_2\.
\zz_2)} \,d^3z_1 \,d^3z_2.
\end{align*}
For harmonium the Wigner distribution factorizes into extracule and
intracule parts:
\begin{equation}
P_\gs(\rr_1,\rr_2;\pp_1,\pp_2) = \frac{1}{\pi^6}
\exp\biggl(-\frac{2H_R}\om \biggr) \exp\biggl( -\frac{2H_r}\mu
\biggr).
\label{eq:west-ham}
\end{equation}
This is reached most efficiently and elegantly by the methods of phase
space quantum mechanics~\cite{Callisto}. One can now obtain $\rho_2$
by the inverse formula of~\eqref{eq:phase-two} ---easily provided for
by Fourier analysis. The pairs density $\rho_2(\rr_1,\rr_2,\rr_1,
\rr_2)$ is recovered by integration over the momenta.

The reduced 1-body phase space (spinless) quasidensity for the ground
state $d_\gs$ is obtained, as in the standard formalism, by
integrating out one set of variables,
\begin{equation}
d_\gs(\rr;\pp) = \frac{2}{\pi^3}
\biggl( \frac{4\om\mu}{(\om + \mu)^2} \biggr)^{3/2}
e^{-2r^2\om\mu/(\om + \mu)} e^{-2p^2/(\om + \mu)}.
\label{eq:fair-ground}
\end{equation}
It is a simple exercise to recover from this Eq.~(2--68) for
$\rho_1(\rr_1,\rr'_1)$ in the treatment in~\cite{Davidson}. The
marginals of $d_\gs$ give the electronic density and momentum density.

It should be recognized that, while $P_\gs$ is a pure state,
mathematically $d_\gs$ describes a mixed state. For Gaussians on phase
space, such as $P_\gs$ and $d_\gs$~too, there are simple rules to
determine whether they represent a pure state~\cite{Robert}, a mixed
state~\cite{Titania}, or neither. Writing $\qq = (\rr_1,\rr_2)$, $\ppp
= (\pp_1,\pp_2)$, $\uu = (\qq,\ppp)$, we find $P_\gs(\uu) = \pi^{-6}
e^{-\uu\.F\uu} = \pi^{-6} e^{-\qq\.A\qq - \ppp\.A^{-1}\ppp}$ where,
amusingly,
\begin{equation}
A = \frac{1}{2} \twobytwo{\om + \mu}{\om - \mu}{\om - \mu}{\om + \mu},
\qquad 
A^{-1} = \frac{1}{2} \twobytwo{\om^{-1} + \mu^{-1}}{\om^{-1} - \mu^{-1}}
{\om^{-1} - \mu^{-1}}{\om^{-1} + \mu^{-1}}.
\label{eq:topsy-turvy} 
\end{equation}
We see that the matrix $F$ corresponding to formula
\eqref{eq:west-ham} is symmetric and symplectic, and therefore
represents indeed a pure state. This is not the case for~$d_\gs$. Thus
our purpose in recovering $P_\gs$ from knowledge of~$d_\gs$
\textit{alone} is akin to putting Humpty Dumpty together again!

\section{\ldots who traced these signs?}
\label{sec:y-escuche}

Since all the relevant quantities factorize, in this section we work
in one dimension for notational simplicity.  Write $u=(r,p)$.  For the
ground state on phase space formula~\eqref{eq:rho-two}
goes~\cite{Pluto} into:
\begin{align}
P_\mathrm{gs}(u_1,u_2;\mathrm{spin}) &= \frac{\bigl( \up_1\dn_2 -
\dn_1\up_2 \bigr) \bigl( \up_{1'}\dn_{2'} - \dn_{1'}\up_{2'})}2
\notag \\
&\; \x \sum_{r,s=0}^\infty c_rc_s\, f_{rs}(u_1) f_{rs}(u_2),
\label{eq:silk-road}
\end{align}
where $c_rc_s = (\pm\sqrt{n_r})(\pm\sqrt{n_s})$, with the signs to be
determined. The $f_{rs}$ are Wigner eigentransitions, the functions on
phase space corresponding to transitions between states. A similar
series for triplet states is established and investigated
in~\cite{Laetitia}.

The explicit form of the $f_{rs}$ will turn out to be well known.  But
first we have to find the good coordinates.  According to a theorem
in~\cite{Titania}, the real quadratic form in the exponent of $d_\gs$
must be symplectically congruent to a diagonal one.  We perform the
transformation
$$
(Q,P) := \bigl((\om\mu)^{1/4}r, (\om\mu)^{-1/4}p\bigr);
\word{or, in shorthand,}  U = Su,
$$
where $S$ is symplectic. Introducing as well the parameter $\la :=
2\sqrt{\om\mu}/(\om + \mu)$, the 1-quasidensity takes the simple form
$$
d_\gs(U) := d_\gs(u(U)) = \frac{2\la}{\pi}\, e^{-\la U^2}.
$$
From now on, we assume in the notation $\dl\ge0$; the treatment for
the ``attractive'' case $\om<\mu$ follows parallel lines. Under that
proviso, we may also write $\la =: \tanh(\b/2)$, so that
\begin{gather*}
\b = \log \frac{1 + \la}{1 - \la}
= 2\log \frac{\sqrt\om + \sqrt\mu}{\sqrt\om - \sqrt\mu},  \word{and}
\sinh(\b/2) = \frac\la{\sqrt{1 - \la^2}}
= \frac{2\sqrt{\om\mu}}{\om - \mu}.
\end{gather*}
{}From the series formula,
$$
\sum_{n=0}^\infty L_n(x)\, e^{-x/2}\, t^n = \frac{1}{(1 - t)}\,
e^{-x(1+t)/2(1-t)},
$$
it follows that
$$
d_\gs(U) = \frac4\pi \, \sinh \frac{\b}{2} \sum_{r=0}^\infty (-1)^r
L_r(2U^2) \, e^{-U^2} e^{-(2r+1)\b/2}.
$$
One recognizes the basis of Wigner eigenfunctions on phase space
corresponding to oscillator states~\cite{Callisto}:
\begin{equation}
f_{rr}(U) = \frac{1}{\pi}\,(-1)^r L_r(2U^2)\, e^{-U^2}
\label{eq:panta-rei};
\end{equation}
thus we realize that $d_\gs$ is in thin disguise a \textit{Gibbs
state}~\cite{Titania} with inverse temperature~$\b$:
\begin{equation}
d_\gs(U) = 4\sinh \frac{\b}{2} \sum_{r=0}^\infty e^{-(2r+1)\b/2}
f_{rr}(U).
\label{eq:high-ground}
\end{equation}
With this we have identified the natural orbitals in the
$U$~variables. Their occupation numbers are
\begin{align}
n_r &= 2\sinh \frac\b2\, e^{-(2r+1)\b/2} = \frac{4\sqrt{\om\mu}}{\om -
\mu}\, \biggl( \frac{\sqrt\om - \sqrt\mu}{\sqrt\om + \sqrt\mu}
\biggr)^{2r+1}
\notag \\
&= \frac{4\sqrt{\om\mu}}{(\sqrt\om + \sqrt\mu\,)^2}\,
\biggl(\frac{\sqrt\om - \sqrt\mu}{\sqrt\om + \sqrt\mu} \biggr)^{2r}.
\label{eq:power-tool}
\end{align}
Notice that $n_0 = 1 - e^{-\b} = Z^{-1}(\b)$, for $Z$ the system's
partition function; also $\sum_r n_r = (1 - e^{-\b}) \sum_r e^{-r\b} =
1$. These $n_r$ have nice square roots:
$$
\sqrt{n_r} = \frac{2(\om\mu)^{1/4}}{\sqrt\om + \sqrt\mu}\,
\biggl( \frac{\sqrt\om - \sqrt\mu}{\sqrt\om + \sqrt\mu} \biggr)^r.
$$
{}Finally, for $r\ge s$ one has~\cite{Callisto}:
$$
f_{rs}(U) := \frac{1}{\pi}\,(-1)^s \sqrt{\frac{s!}{r!}} \,
(2U^2)^{(r-s)/2} e^{-i(r-s)\vth} L_s^{r-s}(2U^2) \, e^{-U^2},
$$
where $\vth := \arctan(P/Q)$. The $f_{sr}$ are complex conjugates
of~$f_{rs}$.

\smallskip

We turn now to the energy functionals, which should allow us to
determine the correct signs in~\eqref{eq:silk-road}. Of course, still
working in dimension one, the total energy in our units is just one
half of the sum of the frequencies; and this is confirmed by the
well-known formula for the Gibbs ensemble (consult~\cite{BartlettM}
for the phase space derivation) represented by~\eqref{eq:high-ground}.
To~wit,
$$
E_\gs = E[d_\gs] = 2\sqrt{\om\mu} \biggl( \frac{1}{e^\b - 1} +
\frac{1}{2} \biggr) = \frac{\om + \mu}{2}.
$$

{}From the viewpoint of DFT, the most interesting part of the energy
is the electron interaction $E_{2_\gs}$. The 1-body Hamiltonian is
given by
$$
h(u) = \frac{p^2}{2} + \frac{\om^2 r^2}{2} = \sqrt{\om\mu}
\biggl(\frac{P^2}{2} + \frac{\om Q^2}{2\mu} \biggr).
$$
It is a simple exercise to obtain the 1-body energy $E_{1\gs}$ by
integrating expression~\eqref{eq:fair-ground} with this observable:
$$
E_{1\gs} = \frac{\om}{2} + \frac{\mu^2 + \om^2}{4\mu}.
$$
It is easily checked that this equals $2\sum_r n_r E_{rr}$, where
$E_{rr}$ denotes the $1$-body energy associated to each natural
orbital in~\eqref{eq:panta-rei}. Thus it must be that
$$
E_{2\gs} = E_\gs - E_{1\gs} = \frac{\mu^2 - \om^2}{4\mu}.
$$

Now we go for the two-body contributions of the natural orbitals. The
interelectronic repulsion potential in~\eqref{eq:Mosh-atom} is
$(\mu^2-\om^2)r_{12}^2/4$, so they are of the form $\sum_{rs} c_r c_s
L_{sr}$, with the $L_{sr}$ given by:
\begin{align*}
L_{sr} &= \frac{\mu^2 - \om^2}{4} \int f_{sr}(q_1;p_1) f_{sr}(q_2;p_2)
(q_1 - q_2)^2 \,dq_1\,dq_2 \,dp_1\,dp_2
\\
&= \frac{\mu^2 - \om^2}{4\sqrt{\om\mu}} \int h_s(Q_1) h_r(Q_1)
(Q_1 - Q_2)^2 h_s(Q_2) h_r(Q_2) \,dQ_1\,dQ_2.
\end{align*}
Here $h_r$ are the usual orthogonal harmonic oscillator eigenfunctions
for unit frequency. We consider the diagonal $r = s$ first, whereby
\begin{align*}
L_{rr} &= \frac{\mu^2 - \om^2}{2\sqrt{\om\mu}} 
\biggl( r + \frac{1}{2} \biggr);  \word{and thus}
\\[\jot]
\sum_r n_r L_{rr} 
&= \frac{\mu^2 - \om^2}{2\sqrt{\om\mu}}\,
\frac{4\sqrt{\om\mu}}{(\sqrt\om + \sqrt\mu\,)^2}\,
\frac{(\om + \mu)(\sqrt\om + \sqrt\mu\,)^2}{16\om\mu}
= \frac{\mu^2 - \om^2}{4\mu}\, \frac{\om + \mu}{2\om} \,.
\end{align*}
We have used that the expected value of $Q^2$ for a harmonic
oscillator eigenstate is $r + \half$ and that the expected value
of~$Q$ is zero. Notice that $\dfrac{\om + \mu}{2\om} < 1$.
We have been considering $\mu<\om$. For attraction between the 
electrons, this contribution changes sign.

A non-vanishing contribution of the off-diagonal part may then come
only from the terms
$$
\pm \frac{\om^2 - \mu^2}{2\sqrt{\om\mu}} \sqrt{n_r n_{r+1}}
\biggl[ \int h_r(Q) h_{r+1}(Q) Q \,dQ \biggr]^2.
$$
We compute:
\begin{align*}
&\frac{\om^2 - \mu^2}{\sqrt{\om\mu}} \sum_{r=0}^\infty \sqrt{n_r
n_{r+1}} \biggl[ \int h_r(Q) h_{r+1}(Q) Q \,dQ \biggr]^2
\\
&\; = \frac{\om^2 - \mu^2}{\sqrt{\om\mu}}\,
\frac{4\sqrt{\om\mu}(\sqrt\om - \sqrt\mu\,)}{(\sqrt\om + \sqrt\mu\,)^3}
\sum_{r=0}^\infty \biggl(
\frac{\sqrt\om - \sqrt\mu}{\sqrt\om + \sqrt\mu} \biggr)^{2r}
\frac{r + 1}{2}
\\
&\; = (\om^2 - \mu^2)\, \frac{2(\sqrt\om - \sqrt\mu\,)}{(\sqrt\om +
\sqrt\mu\,)^3}\, \frac{(\sqrt\om + \sqrt\mu\,)^4}{16\,\om\mu} =
\frac{\om^2 - \mu^2}{4\mu}\, \frac{\om - \mu}{2\om} \,.
\end{align*}
Here we employ $\sum_{r=0}^\infty (r + 1)x^r = (1 - x)^{-2}$. The
factor $(r + 1)/2$ comes from the definition of the emission operators
$a^\7 = (Q - iP)/\sqrt2$ (or the absorption operators), with $a^\7 h_r
= \sqrt{r + 1}\,h_{r+1}$. There is also an overall factor of~$2$
coming from two subdiagonals for each~$r$. Obviously there are no
other contributions.

In conclusion, to minimize the energy we now have to choose
\textit{minus} signs whenever $s = r \pm 1$. This conclusion does not
depend on whether the electrons attract or repel each other. It
determines all the signs in~\eqref{eq:silk-road}. Then the total
2-body energy comes out as
$$
\frac{\mu^2 - \om^2}{4\mu}\, \frac{\om + \mu}{2\om} - \frac{\om^2 -
\mu^2}{4\mu}\, \frac{\om - \mu}{2\om} = \frac{\mu^2 - \om^2}{4\mu},
$$
as it ought to be.

\begin{thm}
The ground state of harmonium is equal to
$$
P_\mathrm{gs}(u_1,u_2) = (-)^{r+s}\sqrt{n_rn_s} \, f_{rs}(u_1)f_{rs}(u_2),
$$
where the $f_{rs}$ are the oscillator eigentransitions up to a
symplectic change of basis, and the $n_r$ have been given
in~\eqref{eq:power-tool}.
\end{thm}

According to the spirit of DFT, the above result has been derived from
knowledge of $d_\mathrm{gs}$ alone; the advantage of working with
Wigner distributions being that the necessary transformations are
completely transparent. However, since we had $P_\mathrm{gs}$
explicitly beforehand, it is nice to make the check.
In~\eqref{eq:silk-road} sum over each subdiagonal, where $r - s = l
\geq 0$:
\begin{align*}
&\sum_{r-s=l}\sqrt{n_r n_s} f_{rs}(u_1) f_{rs}(u_2) =
\frac{n_0}{\pi^2}\, e^{-l\b/2} (2U_1U_2)^l e^{-il(\vth_1 + \vth_2)}
\,e^{-U_1^2-U_2^2}
\\
&\x \sum_{s=0}^\infty \frac{s!}{(l + s)!}\, e^{-s\b}\, L_s^l(2U_1^2)
L_s^l(2U_2^2) = \frac{e^{-(U_1^2 + U_2^2)/\la}}{\pi^2} e^{-il(\vth_1 +
\vth_2)} I_l\biggl( \frac{2U_1U_2}{\sinh(\b/2)} \biggr),
\end{align*}
where $I_l$ is the modified Bessel function. The generating function
identity
$$
I_0(z) + 2 \sum_{l=1}^\infty I_l(z) \cos(l\th) = e^{z\cos\th},
$$
where the sign rule dictates $\th = \vth_1 + \vth_2 + \pi$, yields for
the total sum:
\begin{align}
& \pi^{-2}
e^{-[(U_1^2 + U_2^2)/\la + 2U_1U_2\csch(\b/2)\cos(\vth_1 + \vth_2)]}
\label{eq:semper-sursum}
\\
&\; = \pi^{-2} e^{-\shalf[(r_1^2 + r_2^2)(\om + \mu) + (p_1^2 +
p_2^2)(\om^{-1} + \mu^{-1})]} e^{-r_1r_2(\om - \mu)}
e^{p_1p_2(\mu^{-1} - \om^{-1})},
\notag
\end{align}
which is the correct result. To our knowledge this the first
analytical verification of the two-electron state theorem.

\smallskip

Once the correct sign configuration in the functional is obtained, one
may compare it with standard approximate workhorses like the 
M\"uller functional~\cite{ForgottenCity,FrankLSS}. M\"uller's is a
clever variation on the standard Hartree--Fock functional. For a
two-electron system in the phase space approach, it adopts instead
of~\eqref{eq:silk-road} the form:
\begin{align*}
P_\mathrm{M}(u_1,u_2;\mathrm{spin}) &= \frac{1}{2}\bigl( \up_1\dn_2 -
\dn_1\up_2 \bigr) \bigl( \up_{1'}\dn_{2'} - \dn_{1'}\up_{2'})
\\
&\x \bigg(\sum_{r,s=0}^\infty n_rn_s\, f_{rr}(u_1) f_{ss}(u_2) -
\sum_{r,s=0}^\infty\sqrt{n_rn_s} f_{rs}(u_1) f_{sr}(u_2)\bigg).
\end{align*}
It is well known that for Coulombian
systems the M\"uller functional tends to overbind. The rigorous proof
of this fact for real two-electron atoms given in~\cite{Pluto} is much
more transparent than the one in~\cite{FrankLSS}; it shows that
definite positivity of the Coulomb potential does play a decisive
role, whereas the extra minus signs in M\"uller's functional do not.
For these very reasons the M\"uller functional's tendency to
overcorrelate needs reexamination in harmonium. This is taken up
in~\cite{Laetitia}.

\section{On the correlation functional}
\label{sec:TioGilito}

The Wigner quasiprobability for the best Hartree--Fock state for
harmonium is given by
\begin{align*}
P_\HF(\rr_1,\rr_2;\pp_1,\pp_2)
&= \frac{1}{\pi^6}\, e^{-(r_1^2 + r_2^2)\sqrt{(\om^2 + \mu^2)/2}}
\, e^{-(p_1^2 + p_2^2)/\sqrt{(\om^2 + \mu^2)/2}}
\\
&= \frac{1}{\pi^6}\, e^{-(R^2 + r^2)\sqrt{(\om^2 + \mu^2)/2}}
e^{-(P^2 + p^2)/\sqrt{(\om^2 + \mu^2)/2}}.
\end{align*}
We refer to~\cite{Moshinsky} for the derivation of the Hartree--Fock
wave function for harmonium, from which the above follows
by~\eqref{eq:phase-two}. From the above, comparing with the exact
state under the form~\eqref{eq:west-ham} or~\eqref{eq:semper-sursum},
a ``Moshinsky hole'' in phase space is clearly visible.

The correlation energy $E_c$ is defined as the difference between the
true energy of the electronic system and the Hartree--Fock energy.
For harmonium, use of $P_\HF$ gives
\begin{align*}
&E_\HF = \frac{1}{\pi^6} \int \biggl( \frac{P^2}{2}
+ \frac{\om^2 R^2}{2} + \frac{p^2}{2} + \frac{\mu^2 r^2}{2} \biggr)
\\
&\qquad \x e^{-(R^2 + r^2)\sqrt{(\om^2 + \mu^2)/2}}\,
e^{-(P^2 + p^2)/\sqrt{(\om^2 + \mu^2)/2}}
\,d^3\PP \,d^3\RR \,d^3\pp \,d^3\rr
\\
&\; = \frac{3\sqrt{(\om^2 + \mu^2)/2}}{4}
+ \frac{3\,\om^2}{4\sqrt{(\om^2 + \mu^2)/2}}
+ \frac{3\sqrt{(\om^2 + \mu^2)/2}}{4}
+ \frac{3\mu^2}{4\sqrt{(\om^2 + \mu^2)/2}}
\\
&\; = 3\sqrt{(\om^2 + \mu^2)/2},
\end{align*}
and so the correlation energy is
\begin{align*}
E_c := E_0 - E_\HF = 3 \biggl( \frac{\om + \mu}{2}
- \sqrt{\frac{\om^2 + \mu^2}{2}} \,\biggr)
\sim - \frac{3\,\dl^2}{32\,\om^3},
\end{align*}
as $\dl\downarrow0$ keeping $\om$ fixed, or as $\om\uparrow\infty$
keeping $\dl$ fixed. If we understand the latter as the
``high-density'' limit, the correlation energy vanishes in it.
In~\cite{collant} the high energy limit is understood as the limit
$\mu\uparrow\infty$, with $\om$ kept constant. In both cases the
behaviour of correlation is at variance with that of the Coulomb
potential~\cite{terribletwins}.

As indicated in the introduction, Gill's program for the correlation
energy~\cite{Gillito2,GillitoF} is based exclusively on the Wigner
intracule. Since relative momentum is as important as relative
position in determining interelectronic correlation, it is certainly
appealing to study correlation on phase space. More precisely, the
full Wigner distribution is reduced to the relative variables
$\rr_{12}=\rr_1-\rr_2,\,\pp_{12}=\pp_1-\pp_2$. Then it is conjectured
that the correlation energy is a universal functional of the ``Omega
intracule''~$\Omega$, that is, the intracule on phase space after
reduction to the variables $r_{12},p_{12},w$, with $w$ the angle
between $\rr_{12}$ and $\pp_{12}$, of the form:
$$
E_c = \int_0^\infty\int_0^\infty\int_0^\pi\Omega(r_{12},p_{12},w)
G(r_{12},p_{12},w)\,dw\, dp_{12}\,dr_{12};
$$
or even a functional of the lower intracules:
$$
E_c = \int_0^\infty\int_0^\infty W(r_{12},p_{12})G(r_{12},p_{12})\,
dp_{12}\,dr_{12}; \quad E_c = \int_0^\infty A(s)G(s)\,ds,
$$
with $s:=r_{12}p_{12}$. In our case, recalling $\rr=\rr_{12}/\sqrt2,
\,\pp=\pp_{12}/\sqrt2$, the exact intracule takes the form
$$
\frac1{\pi^3} e^{-\mu r^2 - p^2/\mu} \longrightarrow \frac2\pi
r_{12}^2\,p_{12}^2\,e^{-\mu\,r_{12}^2/2} e^{-p_{12}^2/2\mu} =:
W_\mathrm{ex}(r_{12},p_{12}),
$$
after integration over the angles. While the Hartree-Fock intracule 
is:
$$
\frac2\pi r_{12}^2\,p_{12}^2\,e^{-\sqrt{(\om^2 +
\mu^2)/2}\,r_{12}^2/2} e^{-p_{12}^2/2\sqrt{(\om^2 + \mu^2)/2}} =:
W_\mathrm{HF}(r_{12},p_{12});
$$
which, as observed in~\cite{collant}, is of the same form as the exact
one when replacing~$\dl$ by~$\dl/2$.  As a consequence the ``action
intracules'':
$$
A(s) = \frac2\pi\int_0^\infty r_{12}^2\,\frac{s^2}{r_{12}^2}
\,e^{-\bullet\,r_{12}^2/2} e^{-s^2/2\bullet\,r_{12}^2}\,
\frac{dr_{12}}{r_{12}} = \frac{2s^2}\pi K_0(s)
$$
for any value of~$\bullet$, are precisely the same for the exact and
Hartree-Fock cases. Note that in the exact solution only the intracule
part of the Wigner quasiprobability changes in harmonium, while the
Hartree--Fock method modifies both the extracule and the intracule;
and then both contribute to the correlation energy. However, in
Gill's~$W$-functions the footprints of both harmonium frequencies are
jointly kept.

\section{Conclusion and outlook}
\label{sec:so-and-so}

We have beaten a path from Schmidt series to the exact energy
functional for two-electron systems, and fully determined the latter
for the harmonium model of an atom, by working with Wigner
distributions on phase space. We hope this paper helps to understand
why this kind of approach is bound to grow in importance in the coming
years.

\end{document}